\documentclass[preprint,3p,times,twocolumn]{elsarticle}
\renewcommand*{\today}{March 26, 2019}

\usepackage[caption=false,font=footnotesize,labelfont=sf,textfont=sf]{subfig}
\usepackage{stfloats}
\usepackage{amsmath}
\usepackage{graphicx}
\usepackage{adjustbox}
\usepackage{url}

\usepackage{rotating}

\usepackage{xspace}
\usepackage{paralist}

\newcommand{\eg}{e.\,g.,\xspace}
\newcommand{\ie}{i.\,e.,\xspace}

\newcommand{\cf}{c.\,f.,\xspace}

\begin{document}

\begin{frontmatter}
\title{Feature-Set-Engineering for Detecting Freezing of Gait in Parkinson's Disease using Deep Recurrent Neural Networks}
\author{Spyroula Masiala}
\ead{s.masiala@tilburguniversity.edu}
\author{Willem Huijbers}
\ead{w.huijbers@uvt.nl}
\author{Martin Atzmueller\corref{cor1}}
\ead{m.atzmuller@uvt.nl}

\cortext[cor1]{Corresponding author}
\address{Tilburg University, Department of Cognitive Science and Artificial Intelligence, Warandelaan 2, 5037 AB Tilburg, The Netherlands}

\begin{abstract}
Freezing of gait (FoG) is a common gait disability in Parkinson's disease, that usually appears in its advanced stage. Freeze episodes are associated with falls, injuries, and psychological consequences, negatively affecting the patients' quality of life. For detecting FoG episodes automatically, a highly accurate detection method is necessary.

This paper presents an approach for detecting FoG episodes utilizing a deep recurrent neural network (RNN) on 3D-accelerometer measurements. We investigate suitable features and feature combinations extracted from the sensors' time series data. Specifically, for detecting FoG episodes, we apply a deep RNN with Long Short-Term Memory cells. In our experiments, we perform both user dependent and user independent experiments, to detect freeze episodes.

Our experimental results show that the frequency domain features extracted from the trunk sensor are the most informative feature group in the subject independent method, achieving an average AUC score of 93\%, Specificity of 90\% and Sensitivity of 81\%. 
Moreover, frequency and statistical features of all the sensors are identified as the best single input for the subject dependent method, achieving an average AUC score of 97\%, Specificity of 96\% and Sensitivity of 87\%. Overall, in a comparison to state-of-the-art approaches from literature as baseline methods, our proposed approach outperforms these significantly.
\end{abstract}


\end{frontmatter}


\section{Introduction}
\label{sec:introduction}

Parkinson's disease (PD) is an age-related neurodegenerative disorder that is traditionally marked by motor symptoms, including tremor, muscular rigidity and freezing \cite{moore_freezing_2007,Kalia:2015eh}. 
In the absence of cure, treatment is aimed at alleviating these symptoms \cite{bloem_falls_2004}.
Freezing of gait (FoG), defined as episodes with inability to move, is the most common gait symptom in PD. Almost half of patients suffer from FoG and approximately 28\% experience FoG episodes daily. FoG episodes become more frequent in the advanced stage of PD and they are associated with falls, injuries,
psychological consequences (\eg, \cf~~\cite{nutt_freezing_2011}) and thus negatively affect the patients' quality of life considerably.

Currently, the standard procedure for FoG assessment involves self-reported diaries from patients and clinical assessment in a laboratory.
Most assessment methods are unfortunately subjective and can give biased information. First, FoG questionnaires and diaries depend on the subjective experience and memory of patients. Secondly, patients might exhibit different FoG patterns in laboratory tests than at home. Third, freeze episodes can be difficult to provoke in clinical assessment at a laboratory
or a doctor's office \cite{nieuwboer_cueing_2008}.
Recent studies started to develop wearable assistants to detect FoG episodes \cite{bachlin_wearable_2010,cole_detecting_2011,mazilu_online_2012,mazilu_feature_2013,moore_freezing_2017,delval_objective_2010,djuric-jovicic_quantitative_2014,han_gait_2003,moore_ambulatory_2008,niazmand_freezing_2011,popovic_simple_2010}. These studies  demonstrate, that wearable sensor data (\ie accelerometers) can be used for the automatic detection of FoG episodes in order to help patients suppress freeze episodes. However, for implementation in daily life, a more accurate detection method is necessary.

This paper presents a deep recurrent neural network (RNN) to detect FoG episodes using measurements obtained from 3D-accelerometers placed on the patient's ankle, trunk and thigh. In contrast to other approaches for detecting FoG using deep learning, we apply an RNN with Long Short-Term Memory (LSTM) cells in order to explicitly model and exploit the time series (history) information. 

Specifically, we focus on the methodology of FoG episode detection using a deep RNN and conduct a comprehensive analysis investigating suitable features and feature combinations extracted from the sensors' time-series data. Understanding the important feature sets then enables explication and computational sensemaking~\cite{Atzmueller:DBDBD17,Atzmueller:18:Declare}. Furthermore, we compare the proposed approach to state-of-the-art baseline methods.

\begin{figure*}[!t]
\centering
\caption{Flowchart of the proposed methodology.}\label{fig:frog}
\includegraphics[trim={1cm 0.5cm 1cm 0}, width=2.00\columnwidth]{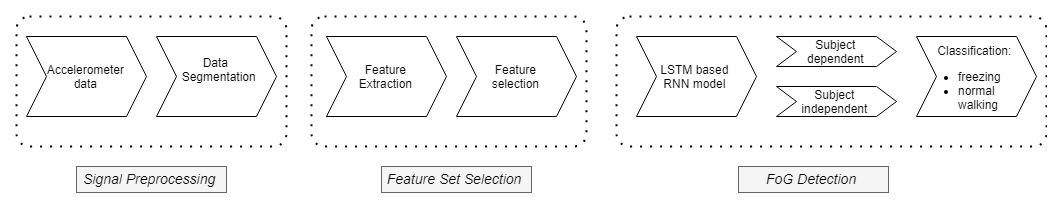}
\end{figure*}

The proposed methodology for FoG episode detection consists of 6 stages depicted in Figure~\ref{fig:frog}: \begin{inparaenum}[(1)] \item Accelerometer data acquisition, \item Data segmentation, \item Feature extraction, \item Feature selection, \item Modeling (subject independent and dependent settings), and \item Classification\end{inparaenum}. Stage one and two can be grouped into a \emph{signal preprocessing phase}, stages three and four into a \emph{feature set selection} phase, while the final \emph{FoG detection} phase is given by stages five to six.

After extracting highly informative features, based on the existing scientific literature, we then apply the proposed approach using a deep RNN with Long Short-Term Memory cells. For evaluation, we apply a de-facto standard benchmark dataset for FoG detection, \ie the Daphnet dataset~\cite{bachlin_wearable_2010}. This dataset has been applied in a multitude of previous approaches, \eg~\cite{bachlin_wearable_2010,cole_detecting_2011,mazilu_online_2012,mazilu_feature_2013,moore_freezing_2017,delval_objective_2010,djuric-jovicic_quantitative_2014,han_gait_2003,moore_ambulatory_2008,niazmand_freezing_2011,popovic_simple_2010}, and as discussed below, we outperform the baseline approaches significantly, in particular for subject-independent evaluation. For replication of our experiments, we make our code for performing the experiments and building the deep RNN model publicly available.\footnote{\url{https://github.com/XXXXX-XXX/deep-fog} (anonymized for blind review)}

It is important to note, that we focus both on user dependent and user independent experiments, to detect freeze episodes, \ie considering episodes for individual patients as well as the overall episodic patterns for our patient group. In total, we perform thirteen experiments with respect to the different sensor placement and we investigated which feature group is enough to successfully detect FoG episodes. Our experimental results show that the frequency domain features extracted from the trunk sensor is the most informative feature group in the subject independent method, achieving an average AUC score of 93\%, Specificity of 90\% and Sensitivity of 81\%. 
Moreover, the frequency and statistical features of all the sensors is identified as the best single input for the subject dependent method, achieving an average AUC score of 97\%, Specificity of 96\% and Sensitivity of 87\%. With our applied LSTM method, our proposed approach outperforms the state-of-the-art baseline methods significantly.

Our contributions are summarized as follows:
\begin{compactenum}
\item We present a methodology for FoG episode detection applying a deep Recurrent Neural Network with Long Short-Term Memory cells exploiting feature sets derived from accelerometer data.
\item We perform a detailed analysis of features (and feature combinations) extracted from the sensors' time series regarding different options for sensor placement.
\item Our experimental results on a real-world benchmark data sets indicate that the trunk sensor yields the most informative feature group. Furthermore, the proposed approach outperforms our baselines, \ie state-of-the-art approaches significantly.
\end{compactenum}

The rest of the paper is organized as follows: Section~\ref{sec:background} outlines the background on FoG episodes, before Section~\ref{sec:related} discusses related work. After that, Section~\ref{sec:method} presents our approach. Next, Section~\ref{sec:results} provides our results, which are discussed in Section~\ref{sec:discussion} in detail. Finally, Section~\ref{sec:conclusions} concludes with a summary and interesting directions for future work.

\section{Background}\label{sec:background}

Several studies showed that wearable assistants utilizing accelerometer data can automatically detect FoG episodes and provide rhythmic auditory cueing that helps the patients to suppress the freeze episodes. Below, we summarize the main concepts for explaining our research context.

\subsection{Overview}
Next to Alzheimer’s disease (AD), Parkinson’s disease (PD) the second most common, age-related neurodegenerative disorder, that effects 1\% of the population over the age of 50 globally. PD progresses slowly many patients live between 10 and 20 years after after diagnosis with a severely reduced quality of life, due to cognitive and motor impairments. The progressive loss of dopaminergic and other neurons impairs motor abilities and causes tremors, bradykinesia and akinesia, commonly known as freezing. In addition to the main motor symptoms, a non-motor symptomatology, for instance personality change, anxiety, dementia, depression, sleep disorder, and hallucinations, may be present as well \cite{jankovic_parkinsons_2008,mhyre_parkinsons_2012}. In the absence of cure, medications or surgery merely target the PD symptoms \cite{bloem_falls_2004}.
While suffering from Parkinson's disease, and given that after some years the above-mentioned symptoms may become not only troublesome but even deadly \cite{factor_parkinsons_2007}, treatment options have been explored in several studies.

Freezing of gait (FoG) is defined as a sudden and transient inability to walk, that affects approximately 50\% of PD patients daily \cite{nutt_freezing_2011,macht_predictors_2007}. 
The disabling phenomena is resistant to the existing parkinsonian medication \cite{bloem_falls_2004}, thus a non-pharmacological treatment has been investigated by numerous studies. 

\subsection{Non-Pharmaceutical Treatment of FoG}
The absence of a complete cure of Parkinson's disease and the inefficacy of pharmacological treatment of FoG incited clinicians, and patients to develop numerous behavioral “tricks”, to overcome freezing episodes, \eg stepping over cracks in the floor, marching to command, shifting body weight, and walking to a beat \cite{rahman_factors_2008}. 

Rhythmic Auditory Stimulation (RAS) forms one of the most efficient instrument in gait improvement among patients suffering from FoG (PWF) \cite{hashimoto_speculation_2006}. 
RAS can provide a regular metronome ticking sound upon the detection of a FoG episode. The sound alerts the patient about the upcoming FoG event, in order to enhance their speed and improve their gait stability \cite{hausdorff_rhythmic_2007}. Recent clinical studies confirmed that the rhythmical ticking sound synchronizes with the gait and helps the PWF suppress the freezing episode and continue walking \cite{donovan_laserlight_2011,suteerawattananon_effects_2004}. 

Nevertheless, despite its safety and efficacy, RAS suffers from a major drawback; its effectiveness weakens through time. Even if it is proven that with the assistance of the rhythmical ticking sound, the duration of a freeze episode is shorter and the patients return to the normal walking patterns \cite{bachlin_wearable_2010}, a long lasting cuing is not recommended \cite{cubo_short-term_2004,nieuwboer_cueing_2008,rubinstein_power_2002} as the efficiency of RAS decrease exponentially through time. Consequently, a context-aware cuing system, that is capable to detect freeze episodes properly and accurately and provide the rhythmical ticking on its onset, is required. 

\section{Related Work}\label{sec:related}

Considering that PD is related to gait disorders and tremor, the movement patterns of such shivers and FoG episodes can be captured by wearable sensors.

Numerous studies on FoG detection use data collected from body sensors, where the 3D-accelerometer sensor appears to be the most popular. Moore et al. \cite{moore_ambulatory_2008} conducted a study in the ambulatory monitoring of FoG, by utilizing 3D-accelerometer data extracted for sensors placed on the left shank of 11 patients suffering from FoG (PWF). They introduced the freezing index (FI) as the power of the body acceleration signal in the freeze band (3 to 8 Hz) divided by the power in the ``locomotor'' band (0.5 to 3 Hz) and they showed that the width of the optimal window was two times the duration of the shortest detected ``freeze'' episode.
Bachlin et al.\cite{bachlin_wearable_2010} developed a real-time wearable device for automatic FOG detection that automatically provides a cueing sound when FOG is detected. They used 3D accelerometer data extracted from body sensors, placed on the ankle, thigh, and trunk of 10 PWF and they reported an average Sensitivity of 73.1\% and Specificity of 81.6\% in the patient-independent method. 
Further studies on FoG detection reported that the system's performance is higher on patient-dependent or group dependent settings \cite{cole_detecting_2011,mazilu_online_2012}, rather than in subject-independent settings \cite{bachlin_wearable_2010,cole_detecting_2011,mazilu_online_2012}.

In 2012, Mazilu et al. developed a wearable FoG detector based on a smart phone, also utilizing the Daphnet dataset \cite{mazilu_online_2012}. They evaluated numerous supervised algorithms, namely Random Forest, C4.5 Decision Trees, Naive Bayes, multilayer perceptron, as well as boasting and bagging methods and they reported that machine learning techniques can adapt successfully the high in dimensionality features of the FoG episodes, instead of the commonly-used manual thresholds. In their most recent work, Mazilu et al. \cite{mazilu_gaitassist:_2014} developed a second daily-life wearable assistant for the PWF, based on a C4.5 Decision Trees algorithm and the FI as its feature.
Similarly, Tripoliti et al.\cite{tripoliti_automatic_2013} designed a FoG detection system, using accelerometer and gyroscope data from PWF and healthy control subjects and evaluated four machine learning algorithms. The Random Forest was reported as the one with the best user-specific performance, with Sensitivity of 81.94\%, Specificity of 98.74\% and Accuracy of 96.11\%.

Cole et al.\cite{cole_detecting_2011} presented a dynamic neural network in their first attempt to detect the FoG episodes in patients with PD, while they performed random activities. They used signals from 3D accelerometer sensors placed on the shin, thigh and forearms and one electromyographic (EMG) sensor placed on the shin of the PD patient. Their FoG detector reported Sensitivity of 83\% and Specificity of 97\%. Hammerla et al. \cite{hammerla_preserving_2013} explored deep recurrent neural networks across the Daphnet dataset \cite{bachlin_wearable_2010} and the FoG detection problem as well. A more recent study by Coste et al. \cite{azevedo_coste_detection_2014} introduced a new method for the observation of gait anomalies and FoG detection. They argued that the FoG criterion (FOGC) -the uninterrupted evaluation of frequency and stride length provides a better indicator of freezing compared to the FI.
Their model obtained results of F1 scores of 76\%. In the same year, Rodriguez et al.\cite{rodriguez-martin_posture_nodate} published a study in which they proposed a posture algorithm to assess 20 PD patients and increased the Specificity of the current FoG detection approaches. Their method increased this metric by about 5\% on average, while maintaining the Sensitivity. They performed their experiments by using a single 3D accelerometer placed on the waist of the subjects and they were able to reduce the false positives when the subjects were not in a standing position; for some patients, the Specificity was increased by 11.9\% preserving the Sensitivity. 
Furthermore, Zhao et al.\cite{zhao_feasibility_2016} used a single accelerometer sensor placed on the back of 23 PD patients who suffered from FoG episodes and they investigated to what extent the FI is able to successfully detect freeze events in terms of Sensitivity and Specificity. However, they targeted only FoG episodes happening for specific walking tasks such as full rapid turns and walking with short steps. Finally, they reported a Sensitivity of 75\% and Specificity of 76\%.

Recently, Rodriguez et al.\cite{rodriguez-martin_home_2017} proposed a new machine learning approach for FoG detection based on a set of 55 features using a Support Vector Machine (SVM) classifier. Their data were composed of signals collected from a single 3D accelerometer sensor placed on the waist of 21 PD patients while they were performing activities of daily life at their home. They evaluated their system under two conditions: (1) a generic model, which was tested using a leave-one-out method and (2) a personalized model, which used part of the dataset of each subject. The reported results provide a Sensitivity and Specificity geometric mean of of 76\% in the generic model and 84\% in the personalized model. However, their method achieved a lower Specificity compared to the most comprehensive FoG detection methods. 

Also, Murad et al. \cite{murad_deep_2017} proposed the use of deep recurrent neural networks (deep RNN) for designing a human activity recognition model, testing their model on the Daphnet dataset~\cite{bachlin_wearable_2010}. They found that the a deep RNN model with one bidirectional layer and two upper unidirectional layers yields the best results in terms of F1 score (93\%). More recently, Camps et al. (2018) \cite{camps_deep_2018} proposed a deep learning method for detecting FoG episodes among PD patients. Their model was based on a convolutional neural network and they evaluated their model on signal data obtained from sensors placed on 21 PD patients. They reported a value of 90\% for the geometric mean between Sensitivity and Specificity.

In contrast to the works summarized above, this paper introduces a novel approach based on a deep recurrent neural network that detects successfully FoG episodes from signal information, not only in subject-dependent settings but in subject-independent as well. For that, we perform a detailed analysis of feature groups of different sensors. We compare our approach and results in relation to the methods presented above as baseline methods on the Daphnet dataset, and show its advantages outperforming the competing approaches significantly.

\section{Method}\label{sec:method}

Below, we first describe the dataset, preprocessing and feature extraction. Next, we describe the proposed approach including the deep recurrent neural network and model architecture. After that, we present the experimental setup, evaluation metrics and the comprehensive analysis of different feature groups within the set of the constructed features. 

\subsection{Dataset Description}
In this paper, we utilize the publicly available Daphnet dataset, which has been developed to benchmark automatic methods to recognize the FoG episodes from wearable 3D accelerometer sensors attached on the leg, the thigh, and the trunk of PD patients \cite{bachlin_wearable_2010}. The Daphnet dataset can be regarded as a publicly available de-facto benchmark dataset for FoG which has been used in multiple studies, \eg~\cite{bachlin_wearable_2010,cole_detecting_2011,mazilu_online_2012,mazilu_feature_2013,moore_freezing_2017,delval_objective_2010,djuric-jovicic_quantitative_2014,han_gait_2003,moore_ambulatory_2008,niazmand_freezing_2011,popovic_simple_2010}.

The Dapnet dataset is provided by the Laboratory for Gait and Neurodynamics at Tel Aviv Sourasky Medical Center (TASMC) and the Wearable Computing Laboratory at ETH Zurich. It contains data collected from three wireless accelerometer sensors placed on the ankle, thigh, and trunk of 10 PD patients. The sensors recorded 3D accelerations at 64Hz and transferred their data to a wearable computer, which was attached to the trunk of the subjects (along with the third sensor) and provided RAS upon the detection of a freeze episode. 

The research protocol was based on two sessions, both designed to replicate a normal daily walking routine. During the first session, the wearable computer collected all the data and conducted online FoG detection, without RAS feedback. In the second session the same procedure was followed, however, the RAS feedback was activated. The subjects performed three basic walking tasks in both 10-15 minutes sessions, more specifically, (1) Walking back and forth in a straight line, including several 180-degrees turns; (2) Random walking with a series of initiated stops and 360 degrees turns; (3) Walking simulating activities of daily living, which included entering and leaving rooms, walking to the kitchen, getting something to drink and returning to the starting room with a cup of water. Afterward, the physiotherapists analyzed the recorded video along with the manual labels to identify the ground truth labels, the duration, the onset, and the end of the FoG episodes as well. The onset of a FoG episode was detected when the locomotion pattern, more specifically the alternation from left to right step, was decelerating, and the end of the FoG episode was considered as the moment that the pattern was accelerating. The physiotherapists labeled manually four activities, specifically standing, walking, turning, and freezing. However, after the video analysis, they categorized the activities standing, walking, and turning as ``no freezing''.

Altogether, 237 FoG episodes were detected (23.7$\pm$20.7 per subject), with duration of between 0.5s and 40.5s (7.3$\pm$6.7s). 93.2\% of the FoG episodes lasted less than 20 seconds, where 50 percent were shorter than 5.4s. 
Consequently, for each 3D accelerometer sensor reading of the Daphnet dataset, the final ground truth class labels are 1 for ``no freeze'' and 2 for ``freeze''. It is worth pointing out that there is an extra class label (annotation) in the Daphnet dataset with the value 0, which is not part of the experiment, for instance the subject war performing activities unrelated to the protocol.

\subsection{Preprocessing}
One of the essential steps in the data mining process is data preprocessing. In our setting, we first deleted annotation 0, because it is not part of the experiment. The next step data preprocessing involves an additional time series, 
$A_{mag}$, which was obtained by computing the magnitude of the three 
accelerations of the signals:
\[A_{mag} = \sqrt{A^2x+A^2y+A^2z}\]
The next fundamental step is the application of windowing techniques, where the sensor signals are sliced into partially overlapping windows. We used a windowing function with a window length of 4 seconds (256 samples) and an overlap of 0.5 seconds (32 samples). After segmentation, relevant features are extracted from each window and the resulting feature vectors are used as training data. 

\subsection{Feature Extraction}
Here, we focused on features described in the literature. We studied two feature extraction schemes, as suggested by Mazilu et al. \cite{mazilu_feature_2013} and Moore et al. \cite{moore_freezing_2017}. For that, we computed statistical, time-domain and frequency-domain features over the signal data within each window. In most studies on FoG detection, a single input is used to extract features; for instance, a single axis from a single sensor (the vertical acceleration (X) of the sensor placed on the ankle) or the calculated magnitude of a single sensor. In our context, we choose both inputs for extracting features, however, we also applied a further method suggested by Moore et al. \cite{moore_freezing_2017}: We use multiple inputs, where the feature values are computed from a matrix of inputs, i.e. multiple axis of multiple body sensors. Overall, we extracted an total number of 145 features for our experiments.

In our first feature extraction scheme, we extract a group of statistical and time-domain features. Time domain features are simple statistical and mathematical metrics, easily computed, and applied to extract basic and significant signal information within the sliced window. In the context of this paper, we extracted the statistical features based on expert knowledge. Mazilu et al., \cite{mazilu_feature_2013} and Moore et al.\cite{moore_freezing_2017} extracted a wide range of statistical and time domain features and investigated their relevance for FoG detection. We decided to extract the average, standard deviation, variance, median, range, maximum and minimum, based on their reported top ranked features. We then extracted the above-mentioned statistical features for each of the three accelerometer axis -x,y,z-, for each body sensor -ankle, thigh and trunk- and for the magnitude of each sensor. Alltogether, we then obtain 84 statistical and time domain features in total.

More recently, Moore et al.\cite{moore_freezing_2017} employed new feature selection techniques based on voting methods with clustering and correlation metrics, to identify the most discriminative ones. They introduced a novel frequency domain feature, the multi-channel FI ($FI_{MC}$), which was ranked as the most informative feature for their anomaly score detector. Similar to the single input FI, the multi-channel FI is calculated from the power of the freeze band (PH) divided by the power of locomotor band (PL), which are summations of single powers over the N axis (N = 3 axis x 3 sensors). 
Also taking into consideration the studies of Mazilu et al.\cite{mazilu_feature_2013} and Moore et al. \cite{moore_freezing_2017} and their reported top-ranked features we therefore extracted the Freeze Index, the Power Index, the energy, the power in the freezing frequency bands, the power in the locomotor frequency bands and the Multi-channel Freeze Index ($FI_{MC}$). 
We extracted the above-mentioned frequency domain features for each of the three accelerometer axis -x,y,z- for each body sensor -ankle, thigh, and trunk- and for the magnitudes of each sensor. 
We therefore obtain 61 additional frequency-domain features in total. A detailed description of the computed statistical and frequency domain features is given in the Table 1. 

\begin{table*}[!] 
\centering
\caption{Detailed description of the statistical and frequency domain features.}
\begin{center}
\begin{tabular}{|p{2cm}|p{5cm}|p{7cm}|}
\hline
\multicolumn{1}{|c|}{\textbf{Feature}} & \multicolumn{1}{c|}{\textbf{Description}} & \multicolumn{1}{c|}{\textbf{Formula}}\\
\hline
Mean & The average signal value over the window & $Mean(\mu_x)=\displaystyle\sum_{i=1}^{\tau}\frac{x_i}{\tau}$\\
\hline
Standard deviation & The mean signal deviation compared to the average signal value over the window & $(\displaystyle\sum_{i=1}^{\tau}\frac{x_i-\mu_x}{\tau})^\frac{1}{2}$\\
\hline
Variance & The square of the standard deviation & $\displaystyle\sum_{i=1}^{\tau}\frac{x_i-\mu_x}{\tau}$\\
\hline
Median & The median signal value over the window & $median_{x_i}(x_i)$\\
\hline
Range & The difference between the maximum and the minimum signal values over the window & $|max_{x_i}(x_i)-min_{x_i}(x_i)|$\\
\hline
Maximum & The maximum signal value over the window & $max_{x_i}(x_i)$\\
\hline
Minimum & The minimum signal value over the window & $min_{x_i}(x_i)$\\
\hline
Energy & The summation of the squared magnitudes of each FFT component of the signal, divided by the 
window length for normalization & $E_x=\displaystyle\sum_{w=0}^{L}MagF_x(w)^2$\\
\hline
Freeze Index & The power in the ``freeze'' band (3-8Hz) divided by the power in the locomotor band (0.5-3Hz) & $FI=\frac{P_H}{P_L}$\\
\hline
Power & The sum of the power in the ``freeze'' band (3-8Hz) divided by the power in the locomotor band (0.5-3Hz) & $Power=P_H+P_L$\\
\hline
Power in the freeze band & The sum of the power spectrum in the ``freeze'' band of frequencies (3-8Hz) divided by the sampling frequency & $P_H=\frac{1}{2f_s}[\displaystyle\sum_{i=H_1+1}^{H_2}[Pxx_n(i)]+\sum_{i=H_1}^{H_2-1}[Pxx_n(i)]$]\\
\hline
Power in the locomotor band & The sum of the power spectrum in the locomotor band of frequencies (0.53Hz) divided by the sampling frequency & $P_L=\frac{1}{2f_s}[\displaystyle\sum_{i=L+1}^{H_1}[Pxx_n(i)]+\displaystyle\sum_{i=L}^{H_1-1}[Pxx_n(i)]$\\
\hline
Multichannel FI ($FI_{MC}$) & The power of the freeze band (PH) divided by the power of locomotor band (PL),that are summations of single powers over the N axis (N = 3 axis x 3 sensors) & $FI_{MC}=\frac{P_H}{P_L}]$ , where 

$P_H=\frac{1}{2f_s}\displaystyle \sum_{n=1}^{N}[\sum_{i=H_1+1}^{H_2}[Pxx_n(i)]+\sum_{i=H_1}^{H_2-1}[Pxx_n(i)]]$ 

$P_L=\frac{1}{2f_s}\displaystyle \sum_{n=1}^{N}[\sum_{i=L+1}^{H_1}[Pxx_n(i)]+\sum_{i=L}^{H_1-1}[Pxx_n(i)]$\\
\hline
\end{tabular}
\end{center}

* Given x is the signal, w the frequency, N the number of inputs, fs the sampling frequency and $\tau$ the window length 

** Given the FFT transform of the signal is computed as: $F_x(w)=\displaystyle\int_{-\infty}^{\infty}{x(t)e^{-jwt}dt}$

*** Given $\overline{F_x(w)}$ the conjugate of the FFT transform of the signal

**** Given power spectrum of a signal is $Pxx_{(w)} = F_x(w) \overline{F_x(w)}$

***** $H_1 =\frac{3N_{FFT}}{f_s}$ , $H_2 =\frac{8N_{FFT}}{f_s}$, $L =\frac{0.5N_{FFT}}{f_s}$
\end{table*}

\subsection{Proposed Approach}

In this paper, we utilize a recurrent neural network based on LSTM cells, in order to exploit the temporal dependencies within the movement data of PD patients.

\subsubsection{Recurrent Neural Network (RNN)}
RNNs are a powerful type of neural network, due to their ``memory'' which allows them to transfer information along the network, for an infinitive time \cite{hochreiter_long_1997}. Specifically, for every element of a sequence, they perform the same task and the output depends on the previous computations. In an RNN, the classification (output) of an input (feature vector) received by a hidden neuron at time $t$ will be guided by all the previous inputs the same hidden neuron received at the time prior to $t$. During the past 30 years, LSTM networks have proved successful at a range of tasks requiring long memory, including music generation \cite{eck_first_2002}, handwriting recognition \cite{graves_supervised_2012,liwicki_novel_2007}, and speech recognition \cite{graves_framewise_2005,graves_connectionist_2006}. The main reason for the success is that LSTM networks use a long range of contextual information. For our study, we will utilize an LSTM based RNN since a human activity recognition task is a classical sequence analysis problem, suitable for an LSTM based recurrent neural network.

\subsubsection{Model Architecture}
LSTM-based RNNs with deeper architecture are built by stacking multiple LSTM layers. The depth of neural networks is generally associated with the success of the approach on a wide range of challenging prediction problems. Given that LSTMs operate on sequence data, it means that the addition of layers adds levels of abstraction of input observations over time. In effect, chunking observations over time or representing the problem at different time scales. Similar to the neural networks with deep architectures, multilayer LSTM models have been successfully used in speech recognition \cite{graves_hybrid_2013,graves_speech_2013,zhang_highway_2015}, as the experimental results suggest that this type of recurrent neural network outperforms the normal LSTM networks. Similarly, for human activity recognition tasks, the deep LSTM based RNN model outperforms the normal LSTM models \cite{inoue_deep_2016}. Stacked LSTMs are now a stable technique for challenging sequence prediction problems. A Stacked LSTM architecture can be defined as an LSTM model comprised of multiple LSTM layers. 

\begin{figure*}[!t]
\centering
\caption{The structure of our LSTM based RNN model consisting of an input layer which receives the 3D input (feature pool), two hidden layers containing LSTM cells and a final dense layer with a sigmoid activation function which return the output (freeze or non-freeze).}\label{fig:lstm}
\includegraphics[trim={1.1cm 0cm 1.1cm 0}, width=2.00 \columnwidth]{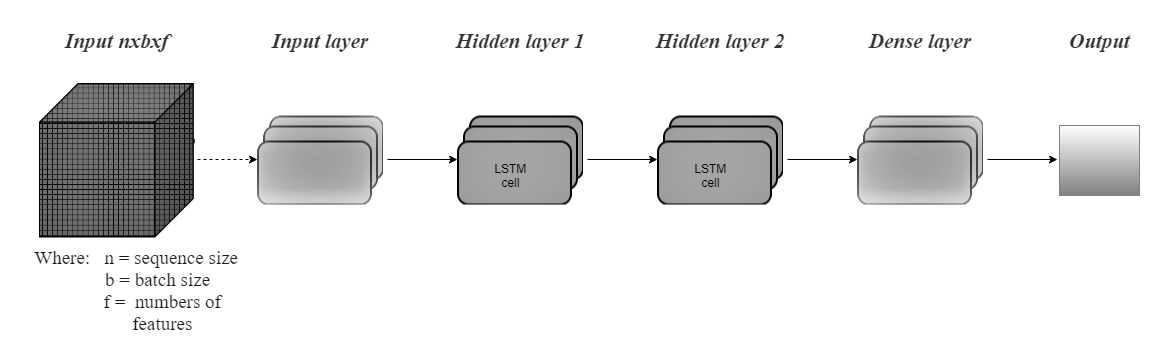}
\end{figure*}

In order to form a model capable of learning richer data representations, we build a Stacked LSTM recurrent neural network, with two LSTM layers (see Figure~\ref{fig:lstm}). Given the 3D accelerometer data, collected from three sensors placed on user's body, we use a sliding window with a length of four seconds; we extract features from each window and we utilize them as input sequences for the LSTM model. For an RNN model, the input involves three dimensions instead of two like the most machine learning algorithms. The three dimensions are the sequence size, the batch size, and the number of features. In each experiment of this research, the single input of our LSTM-based RNN model will depend on the group of the extracted features we choose to feed the network. Thus, the input layer receives the 3D volumetric input. After receiving the input values, the input layer connects into the first LSTM recurrent layer with several memory blocks (``smart'' neurons), that afterward connects into the second LSTM recurrent layer. Finally, since we treat FoG prediction as a binary classification problem, the second LSTM layer, feeds into a fully connected dense layer with a sigmoid activation function, which obtains the predictions. 

\subsection{Experiments and Evaluation}
We evaluated different feature groups extracted from different sensor placements in terms of detection accuracy in user-dependent and user-independent experiments. The reference for all our evaluations is the ground truth annotation provided by physiotherapists in the Daphnet dataset.

The performance of our proposed approach is based on window evaluation, i.e. for each window the output is compared to the reference annotation. The windows that are correctly labeled as ``freeze'' episodes are counted as True Positive (TP), while the wrongly labeled as FoG episode are counted as False Positives (FP). The windows that the system failed to correctly label as a FoG episode are counted as False Negatives (FN) and the windows correctly labeled as no FoG are counted as True Negatives (TN). The Sensitivity \((Sens=\frac{TP}{TP+FN})\) measures the ratio of the correctly labeled FoG windows to the number of the referenced FoG windows, while the Specificity \((Spec=\frac{TN}{TN+FP})\) calculates the ratio of correctly predicted no-FoG windows to the number of the referenced no-FoG windows. 

Additionally, the area under the curve (AUC) in ROC space is reported as performance metrics to evaluate our predictive model. An ROC curve plots the True Positive Rate (on the y-axis) versus the False Positive Rate (on the x-axis) for every possible classification threshold. Specifically, the True Positive Rate represents the ratio of the correctly labeled as ``freeze'' episodes to the number of actual ``freeze'' episodes, while the False Positive Rate represents the ration of the wrong predicted as ``freeze'' episodes to the number of the actual ``non-freeze'' episodes. The AUC is used to quantify the performance of a classifier, estimating the percentage of the area under the ROC curve, where an AUC score equal to 0.5 indicates a purely random classifier, while an AUC score equal to 1 indicates the perfect classifier. 

We perform several experiments and in each experiment, we used different feature groups. More specifically, we performed 3 experiments, each one of them consist of 4 rounds. The number of the experiment indicate a different feature group, i.e. statistical features for the first one, frequency features for the second one and a summation of statistical and frequency features for the third one. The number of each round indicates the different body sensor, i.e. ankle for the first one, thigh for the second, trunk for the third and a summation of them for the fourth one. Moreover, we perform a fourth experiment, where we use as input for our model the 25 most informative features which we determine by recursive feature elimination.

\subsection{Environment}
Our experiments were executed in a computing environment with the following features: Intel Core i7-6500U Processor (8 cores at 2.5 GHz), with 16GB of memory. The code for training the deep learning models implemented was written in Python (version 3.6), using the Keras library (version 2.0.6) \cite{chollet2015keras} running on top of TensorFlow (version tensorflow-gpu 1.7.0). Furthermore, the code implemented for this work is publicly available at \url{https://github.com/XXXXX-XXX/deep-fog}.\footnote{Anonymized for blind review -- this will be made available for the final version.}

\section{Results}\label{sec:results}

For evaluation, we considered both subject independent as well as subject dependent settings, accordingly using different foci for constructing the respective training and test datasets.

\subsection{Subject independent method}
In constrast to several computing approaches, our proposed RNN-based approach was also evaluated in user-independent settings. 
The RNN model was thus trained on data obtained from five participants, specifically patients 5, 6,7,8,9. Afterwards, it was tested on data from three remaining participants, specifically patients 1, 2 and 3. In order too obtain a more balanced dataset, we used the Synthetic Minority Oversampling Technique (SMOTE) on the training set. Afterwards, we trained our RNN model on the new, balanced training set, and we evaluated the performance of our model on the unseen test data obtained from the rest 3 patients. 

The results of the experiments depicted in Table~\ref{Table:2} show that the best performing model in the subject independent setting is the one where we use frequency domain data from the trunk sensor as input for our model. Specifically, our RNN model achieves an AUC score of 93\%, a Specificity of 85\% and a Sensitivity of 89\%. In other words, our model is able to detect 85 non-FoG episodes out of 100 non-FoG episodes and 89 FoG episodes out of 100 FoG episodes. The performance of our model is thus rather accurate and also very satisfying for practical settings since it can not only detect the ``freeze'' events (FoG episodes), but also the non-freeze events so as not to disturb the patients with unnecessary rhythmic sounds in an applied setting.

\begin{table*}[!t]
\centering
\caption{Average AUC, Specificity and Sensitivity for our LSTM based RNN model in the best performing experiments}\label{Table:2}
%
%
\begin{tabular}{|l|l|l|l|}
\hline
\multicolumn{1}{|c|}{\textbf{Experiment}} & \multicolumn{1}{c|}{\textbf{AUC score}} & \multicolumn{1}{c|}{\textbf{Specificity}} & \multicolumn{1}{c|}{\textbf{Sensitivity}} \\ \hline
Frequency features, ankle sensor & 93\% & 90\% & 81\% \\
\hline
Frequency features, trunk sensor & 93\% & 85\% & 89\% \\
\hline
Frequency and statistical features, ankle sensor & 92\% & 86\% & 84\% \\
\hline
Frequency and statistical features, thigh sensor & 89\% & 82\% & 80\% \\
\hline
Frequency and statistical features, all the sensors & 87\% & 80\% & 78\% \\
\hline
\end{tabular}
\end{table*}

\subsection{Subject dependent method}

In the subject-dependent experiments, both training and test datasets are obtained from the same respective patient from the Daphnet dataset. We thus divide the data from each patient in a balanced way, 70\% of the instances create the training dataset and 30\% of the instances the test dataset. We preprocessed the training dataset using the Synthetic Minority Oversampling Technique (SMOTE) in an attempt to overcome the skewed imbalance and we get a new over-sampled training dataset where both ``freeze'' and ``non-freeze'' classes were equally represented. In the user-dependent method, we train our RNN model on the training set of each patient and evaluate the performance of our neural network model on the test dataset (unseen data) of the same patient. 
We report results on average performance measures on the whole dataset for our LSTM based RNN model. 
According to the results presented in Table~\ref{Table:3} the best performances were obtained by statistical and time domain data extracted from the signals of the trunk sensor as single input for our model. Specifically, in the subject dependent method, our model obtained an 95\% AUC score, with 87\% Specificity and 83\% Sensitivity values. 
\begin{table*}[t]
\centering
\caption{Average AUC, Specificity and Sensitivity for our LSTM based RNN model in the best performing experiments}\label{Table:3}
%
%
\begin{tabular}{|l|l|l|l|}
\hline
\multicolumn{1}{|c|}{\textbf{Experiment}} & \multicolumn{1}{c|}{\textbf{AUC score}} & \multicolumn{1}{c|}{\textbf{Specificity}} & \multicolumn{1}{c|}{\textbf{Sensitivity}} \\ \hline
Statistical features, trunk sensor & 95\% & 99\% & 83\% \\
\hline
Frequency and statistical features, all the sensors & 97\% & 96\% & 87\% \\
\hline
Frequency and statistical features, thigh sensor & 96\% & 94\% & 82\% \\
\hline
Frequency and statistical features, ankle sensor & 96\% & 94\% & 82\% \\
\hline
Frequency and statistical features, trunk sensor & 96\% & 96\% & 81\% \\
\hline
\end{tabular}
\end{table*}

\section{Discussion}\label{sec:discussion}
In this paper, we investigated the performance of a deep learning approach for FoG detection using wearable sensors. We applied a deep RNN using LSTM memory cells. In our experiments we observed the performance of our proposed approach using new features (Moore et al. \cite{moore_freezing_2017}) that are more relevant and informative than those previously employed in FoG detection studies.

Our FoG detection model achieves an AUC score of 93\%, with a Specificity of 85\% in the subject independent method (frequency domain features extracted from the signals of the trunk sensor). Moreover, our proposed approach also works well in user dependent settings as well, reporting an average AUC score of 95\%, with Specificity of 87\% and Sensitivity of 83\% (statistical and time domain data extracted from the signals of the trunk sensor). Our results also show that the frequency domain features extracted from the trunk sensor, especially the FI, are the most informative, a finding that confirms the results of Moore et al.\cite{moore_freezing_2017}.

In relation to previous work, we can use the state-of-the-art methods as baselines for comparison with our presented approach, since we use the same de-facto benchmark dataset (\ie the Daphnet dataset) for FoG detection. To summarize, the results previously obtained by Bachlin et al. \cite{bachlin_wearable_2010}, Hammerla et al. \cite{hammerla_preserving_2013}, Mazilu et al.\cite{mazilu_online_2012} and Moore et al.\cite{moore_freezing_2017} were quite lower than the obtained by the presented model, at least in terms of Sensitivity in the subject independent method.

In particular, Bachlin et al. \cite{bachlin_wearable_2010} achieved a Sensitivity
87.1\% and a Specificity of
86.9\% which is significantly lower than the scores obtained by our proposed approach. Furthermore, ompared to the results of Hammerla et al. \cite{hammerla_preserving_2013}, both Specificity and Sensitivity (82\%) were also significantly lower than the ones reported in this work. In a similar way, Moore et al. \cite{moore_freezing_2017} reported a significantly lower level of Specificity (84.5\%) and Sensitivity (87.5\%) compared to our results. However, while Mazilu et al. \cite{mazilu_online_2012} achieved a higher level of Speciﬁcity (95.38\%), they were only able to cope with a quite lower value of Sensitivity (66.25\%).

Moreover, the performance of our model in the subject dependent method is comparable as well. Bachlin et al.\cite{bachlin_wearable_2010} reported a lower level Specificity (92.4\%), however they achieved a quite higher Sensitivity of 88.6\%. Compared to the results of Mazilu et al. \cite{mazilu_online_2012} both Sensitivity (99.69\%) and Specificity (99.96\%) are higher than the ones reported in this work. However, as also noted by the authors~\cite{mazilu_online_2012} there is a bias in their experiments concerning their experimental setup, and it is highly likely that these high values are due to overfitting to the majority class in their evaluation procedure. This is due to the fact that they applied purely random 10-fold cross validation, in contrast to our ``balanced'' evaluation procedure using SMOTE. Therefore, besides our superior performance in AUC (95\%), in our proposed approach the average Sensitivity score was always higher than 55\% and the average Specificity score was always higher than 80\% for each patient.

At this point, we would also like to draw attention to the fact that Bachlin et al. \cite{bachlin_wearable_2010} reported significant worse result in terms of Speciﬁcity for the patient 01 (38.7\% Speciﬁcity and 97.1\% sensitivity) and in terms of Sensitivity for the patient 08 (28.7\% sensitivity and 87.7\% speciﬁcity). In our experiments, the high variability in the movement performance of the patients reported in previous studies, is absent. 
The analysis of different sensor locations showed that the trunk sensor is sufficient for the successful detection of FoG episodes and the promising results point out the emergence of a smartphone or a wearable assistant that can automatically detect freeze episodes and provide RAS to the patients. 

Finally, regarding the model design, our proposed approach significantly outperforms current approaches in subject-independent settings (offline FoG detection) in terms of Sensitivity, to the best of the authors' knowledge. Previous results obtained by \cite{bachlin_wearable_2010,mazilu_online_2012,hammerla_preserving_2013,moore_freezing_2017} demonstrated significantly lower performance than the results obtained from our presented approach. As reported in 
Table~\ref{Table:4}, Mazilu et al. \cite{mazilu_online_2012} and Bachlin et al. \cite{bachlin_wearable_2010} achieved a higher Specificity but with a lower Sensitivity. In our experiments, significance was assessed adapting the method proposed by Demsar et al.~\cite{demsar2006statistical}.

\begin{table*}[t]
\centering
\caption{ Comparison of our proposed model against previous methods in terms of FoG detection}\label{Table:4}
\begin{tabular}{|l|l|l|l|}
\hline
\multicolumn{1}{|c|}{\textbf{Reference }} & \multicolumn{1}{c|}{\textbf{Window size (Tolerance)}} & \multicolumn{1}{c|}{\textbf{Specificity}} & \multicolumn{1}{c|}{\textbf{Sensitivity}} \\ \hline
Bachlin et al.\cite{bachlin_wearable_2010} & 4s (1s) & 81.6\% (online) 86.9\% (offline) & 73.1\% (online) 87.1\% (offline \\
\hline
Hammerla et al. \cite{hammerla_preserving_2013} & - & 82\% & 82\% \\
\hline
S. Mazilu et al. \cite{mazilu_online_2012} & 4s (1s) & 95.38\% & 66.25\% \\
\hline
S. T. Moore et al. \cite{moore_freezing_2017} & 8s (0.5s) & 84.5\% & 87.5\% \\
\hline
Our model & 4s (0.5s) & 85\% & 89\% \\
\hline
\end{tabular}
\end{table*}

\section{Conclusion}\label{sec:conclusions}

This paper presented a deep recurrent neural network to detect FoG episodes using 3D-acceleration measurements obtained from sensors placed on the patient's ankle, trunk and thigh.
In contrast to other approaches for detecting FoG using deep learning, we proposed a methodology for FoG episode detection applying a Recurrent Neural Network with Long Short-Term Memory cells exploiting specific feature sets derived from accelerometer data. In particular, we investigated suitable features and feature combinations extracted from the sensors' time-series data. Here, we performed a comprehensive and detailed analysis of features (and feature combinations) extracted from the sensors' time series regarding different options for sensor placement. Understanding the important feature sets then enables explication and computational sensemaking~\cite{Atzmueller:DBDBD17,Atzmueller:18:Declare}, for example, for understanding the impact of single features, or enhanced profiling of FoG episodes in specific subgroups, \eg~\cite{APB:05b,valmarska17}.

In summary, our experimental results on a real-world benchmark data sets indicated that the trunk sensor yields the most informative feature group. Furthermore, the proposed approach outperforms a set of state-of-the-art approaches selected as baselines significantly, in particular for the subject-independent method. This is especially important towards practical application.

For future work, we aim to incorporate more complex feature models, \eg in order to include dependencies between sensors. Here, also the inclusion of a priori knowledge in knowledge-based approaches, \eg~\cite{ABP:06a,PABHLB:08} and the combination with deep learning techniques is a promising direction~\cite{battaglia2018relational}. Then, both the model construction itself can be complemented, as well as its explicative capabilities~\cite{Atzmueller:DBDBD17,Atzmueller:18:Declare}, \eg regarding the transparency of the model, its interpretability and the generation of explanations. Furthermore, we aim to explore the applicability of the proposed approach and model in enhanced experiments on further datasets, also in similar domains like explicative activity recognition~\cite{AHTK:18}, and its practical implementation in online settings for FoG detection.


\begin{thebibliography}{10}
\expandafter\ifx\csname url\endcsname\relax
  \def\url#1{\texttt{#1}}\fi
\expandafter\ifx\csname urlprefix\endcsname\relax\def\urlprefix{URL }\fi
\expandafter\ifx\csname href\endcsname\relax
  \def\href#1#2{#2} \def\path#1{#1}\fi

\bibitem{moore_freezing_2007}
O.~Moore, C.~Peretz, N.~Giladi, Freezing of gait affects quality of life of
  peoples with {Parkinson}'s disease beyond its relationships with mobility and
  gait, Mov. Disord. 22~(15) (2007) 2192--2195.
\newblock \href {http://dx.doi.org/10.1002/mds.21659}
  {\path{doi:10.1002/mds.21659}}.

\bibitem{Kalia:2015eh}
L.~V. Kalia, A.~E. Lang, {Parkinson's disease.}, Lancet 386~(9996) (2015)
  896--912.
\newblock \href {http://dx.doi.org/10.1016/S0140-6736(14)61393-3}
  {\path{doi:10.1016/S0140-6736(14)61393-3}}.

\bibitem{bloem_falls_2004}
B.~R. Bloem, J.~M. Hausdorff, J.~E. Visser, N.~Giladi, Falls and freezing of
  gait in {Parkinson}'s disease: a review of two interconnected, episodic
  phenomena, Mov. Disord. 19~(8) (2004) 871--884.
\newblock \href {http://dx.doi.org/10.1002/mds.20115}
  {\path{doi:10.1002/mds.20115}}.

\bibitem{nutt_freezing_2011}
J.~G. Nutt, B.~R. Bloem, N.~Giladi, M.~Hallett, F.~B. Horak, A.~Nieuwboer,
  Freezing of gait: moving forward on a mysterious clinical phenomenon, Lancet
  Neurol 10~(8) (2011) 734--744.
\newblock \href {http://dx.doi.org/10.1016/S1474-4422(11)70143-0}
  {\path{doi:10.1016/S1474-4422(11)70143-0}}.

\bibitem{nieuwboer_cueing_2008}
A.~Nieuwboer, Cueing for freezing of gait in patients with {Parkinson}'s
  disease: a rehabilitation perspective, Mov. Disord. 23 Suppl 2 (2008)
  S475--481.
\newblock \href {http://dx.doi.org/10.1002/mds.21978}
  {\path{doi:10.1002/mds.21978}}.

\bibitem{bachlin_wearable_2010}
M.~Bachlin, M.~Plotnik, D.~Roggen, I.~Maidan, J.~M. Hausdorff, N.~Giladi,
  G.~Troster, Wearable {Assistant} for {Parkinsons} {Disease} {Patients} {With}
  the {Freezing} of {Gait} {Symptom}, IEEE Transactions on Information
  Technology in Biomedicine 14~(2) (2010) 436--446.
\newblock \href {http://dx.doi.org/10.1109/TITB.2009.2036165}
  {\path{doi:10.1109/TITB.2009.2036165}}.

\bibitem{cole_detecting_2011}
B.~T. Cole, S.~H. Roy, S.~H. Nawab, Detecting freezing-of-gait during
  unscripted and unconstrained activity, Conf Proc IEEE Eng Med Biol Soc 2011
  (2011) 5649--5652.
\newblock \href {http://dx.doi.org/10.1109/IEMBS.2011.6091367}
  {\path{doi:10.1109/IEMBS.2011.6091367}}.

\bibitem{mazilu_online_2012}
S.~Mazilu, M.~Hardegger, Z.~Zhu, D.~Roggen, G.~TrÃ¶ster, M.~Plotnik, J.~M.
  Hausdorff, Online detection of freezing of gait with smartphones and machine
  learning techniques, in: {International} {Conference} on {Pervasive}
  {Computing} {Technologies} for {Healthcare}, 2012, pp. 123--130.
\newblock \href {http://dx.doi.org/10.4108/icst.pervasivehealth.2012.248680}
  {\path{doi:10.4108/icst.pervasivehealth.2012.248680}}.

\bibitem{mazilu_feature_2013}
S.~Mazilu, A.~Calatroni, E.~Gazit, D.~Roggen, J.~M. Hausdorff, G.~TrÃ¶ster,
  Feature {Learning} for {Detection} and {Prediction} of {Freezing} of {Gait}
  in {Parkinson}â€™s {Disease}, Springer, Berlin, 2013, pp. 144--158.

\bibitem{moore_freezing_2017}
S.~T. Moore, T.~T. Pham, S.~J.~G. Lewis, D.~N. Nguyen, E.~Dutkiewicz, A.~J.
  Fuglevand, A.~L. McEwan, P.~H.~W. Leong, Freezing of {Gait} {Detection} in
  {Parkinson}â€™s {Disease}: {A} {Subject}-{Independent} {Detector} {Using}
  {Anomaly} {Scores}.

\bibitem{delval_objective_2010}
A.~Delval, A.~H. Snijders, V.~Weerdesteyn, J.~E. Duysens, L.~Defebvre,
  N.~Giladi, B.~R. Bloem, Objective detection of subtle freezing of gait
  episodes in {Parkinson}'s disease, Mov. Disord. 25~(11) (2010) 1684--1693.
\newblock \href {http://dx.doi.org/10.1002/mds.23159}
  {\path{doi:10.1002/mds.23159}}.

\bibitem{djuric-jovicic_quantitative_2014}
M.~Djuric-Jovicic, N.~Jovicic, S.~RadovanoviÄ‡, N.~Kresojevic, V.~KostiÄ‡,
  M.~PopoviÄ‡, Quantitative and qualitative gait assessments in {Parkinson}'s
  disease patients, Military Medical and Pharmaceutical Journal of Serbia 71.

\bibitem{han_gait_2003}
J.~H. Han, W.~J. Lee, T.~B. Ahn, B.~S. Jeon, K.~S. Park, Gait analysis for
  freezing detection in patients with movement disorder using three dimensional
  acceleration system, in: Proc. {Annual} {International} {Conference} of the
  {IEEE} {Engineering} in {Medicine} and {Biology} {Society}, Vol.~2, 2003, pp.
  1863--1865 Vol.2.
\newblock \href {http://dx.doi.org/10.1109/IEMBS.2003.1279781}
  {\path{doi:10.1109/IEMBS.2003.1279781}}.

\bibitem{moore_ambulatory_2008}
S.~T. Moore, H.~G. MacDougall, W.~G. Ondo, Ambulatory monitoring of freezing of
  gait in {Parkinson}'s disease, J. Neurosci. Methods 167~(2) (2008) 340--348.
\newblock \href {http://dx.doi.org/10.1016/j.jneumeth.2007.08.023}
  {\path{doi:10.1016/j.jneumeth.2007.08.023}}.

\bibitem{niazmand_freezing_2011}
K.~Niazmand, K.~Tonn, Y.~Zhao, U.~M. Fietzek, F.~Schroeteler, K.~Ziegler, A.~O.
  Ceballos-Baumann, T.~C. Lueth, Freezing of {Gait} detection in {Parkinson}'s
  disease using accelerometer based smart clothes, in: 2011 {IEEE} {Biomedical}
  {Circuits} and {Systems} {Conference} ({BioCAS}), 2011, pp. 201--204.
\newblock \href {http://dx.doi.org/10.1109/BioCAS.2011.6107762}
  {\path{doi:10.1109/BioCAS.2011.6107762}}.

\bibitem{popovic_simple_2010}
M.~B. Popovic, M.~Djuric-Jovicic, S.~Radovanovic, I.~Petrovic, V.~Kostic, A
  simple method to assess freezing of gait in {Parkinson}'s disease patients,
  Brazilian Journal of Medical and Biological Research 43~(9) (2010) 883--889.
\newblock \href {http://dx.doi.org/10.1590/S0100-879X2010007500077}
  {\path{doi:10.1590/S0100-879X2010007500077}}.

\bibitem{Atzmueller:DBDBD17}
M.~Atzmueller, {Onto Explicative Data Mining: Exploratory, Interpretable and
  Explainable Analysis}, in: Proc. Dutch-Belgian Database Day, TU Eindhoven,
  Netherlands, 2017.

\bibitem{Atzmueller:18:Declare}
M.~Atzmueller, {Declarative Aspects in Explicative Data Mining for
  Computational Sensemaking}, in: Proc. International Conference on Declarative
  Programming (DECLARE), Springer, Heidelberg, Germany, 2018.

\bibitem{jankovic_parkinsons_2008}
J.~Jankovic, Parkinson's disease: clinical features and diagnosis, J. Neurol.
  Neurosurg. Psychiatr. 79~(4) (2008) 368--376.
\newblock \href {http://dx.doi.org/10.1136/jnnp.2007.131045}
  {\path{doi:10.1136/jnnp.2007.131045}}.

\bibitem{mhyre_parkinsons_2012}
T.~R. Mhyre, J.~T. Boyd, R.~W. Hamill, K.~A. Maguire-Zeiss, Parkinsonâ€™s
  {Disease}, Subcell Biochem 65 (2012) 389--455.

\bibitem{factor_parkinsons_2007}
S.~A. Factor, W.~Weiner, Parkinson's {Disease}: {Diagnosis} \& {Clinical}
  {Management}, {Second} {Edition}, Demos Medical Publishing, 2007.

\bibitem{macht_predictors_2007}
M.~Macht, Y.~Kaussner, J.~C. MÃ¶ller, K.~Stiasny-Kolster, K.~M. Eggert, H.-P.
  KrÃ¼ger, H.~Ellgring, Predictors of freezing in {Parkinson}'s disease: a
  survey of 6,620 patients, Mov. Disord. 22~(7) (2007) 953--956.
\newblock \href {http://dx.doi.org/10.1002/mds.21458}
  {\path{doi:10.1002/mds.21458}}.

\bibitem{rahman_factors_2008}
S.~Rahman, H.~J. Griffin, N.~P. Quinn, M.~Jahanshahi, The factors that induce
  or overcome freezing of gait in {Parkinson}'s disease, Behav Neurol 19~(3)
  (2008) 127--136.

\bibitem{hashimoto_speculation_2006}
T.~Hashimoto, Speculation on the responsible sites and pathophysiology of
  freezing of gait, Parkinsonism \& Related Disorders 12~(Supplement 2) (2006)
  S55--S62.
\newblock \href {http://dx.doi.org/10.1016/j.parkreldis.2006.05.017}
  {\path{doi:10.1016/j.parkreldis.2006.05.017}}.

\bibitem{hausdorff_rhythmic_2007}
J.~M. Hausdorff, J.~Lowenthal, T.~Herman, L.~Gruendlinger, C.~Peretz,
  N.~Giladi, Rhythmic auditory stimulation modulates gait variability in
  {Parkinson}'s disease, Eur. J. Neurosci. 26~(8) (2007) 2369--2375.
\newblock \href {http://dx.doi.org/10.1111/j.1460-9568.2007.05810.x}
  {\path{doi:10.1111/j.1460-9568.2007.05810.x}}.

\bibitem{donovan_laserlight_2011}
S.~Donovan, C.~Lim, N.~Diaz, N.~Browner, P.~Rose, L.~R. Sudarsky, D.~Tarsy,
  S.~Fahn, D.~K. Simon, Laserlight cues for gait freezing in {Parkinson}'s
  disease: an open-label study, Parkinsonism Relat. Disord. 17~(4) (2011)
  240--245.
\newblock \href {http://dx.doi.org/10.1016/j.parkreldis.2010.08.010}
  {\path{doi:10.1016/j.parkreldis.2010.08.010}}.

\bibitem{suteerawattananon_effects_2004}
M.~Suteerawattananon, G.~S. Morris, B.~R. Etnyre, J.~Jankovic, E.~J. Protas,
  Effects of visual and auditory cues on gait in individuals with {Parkinson}'s
  disease, J. Neurol. Sci. 219~(1-2) (2004) 63--69.
\newblock \href {http://dx.doi.org/10.1016/j.jns.2003.12.007}
  {\path{doi:10.1016/j.jns.2003.12.007}}.

\bibitem{cubo_short-term_2004}
E.~Cubo, S.~Leurgans, C.~G. Goetz, Short-term and practice effects of metronome
  pacing in {Parkinson}'s disease patients with gait freezing while in the 'on'
  state: randomized single blind evaluation, Parkinsonism Relat. Disord. 10~(8)
  (2004) 507--510.
\newblock \href {http://dx.doi.org/10.1016/j.parkreldis.2004.05.001}
  {\path{doi:10.1016/j.parkreldis.2004.05.001}}.

\bibitem{rubinstein_power_2002}
T.~C. Rubinstein, N.~Giladi, J.~M. Hausdorff, The power of cueing to circumvent
  dopamine deficits: a review of physical therapy treatment of gait
  disturbances in {Parkinson}'s disease, Mov. Disord. 17~(6) (2002) 1148--1160.
\newblock \href {http://dx.doi.org/10.1002/mds.10259}
  {\path{doi:10.1002/mds.10259}}.

\bibitem{mazilu_gaitassist:_2014}
S.~Mazilu, U.~Blanke, M.~Hardegger, G.~TrÃ¶ster, E.~Gazit, M.~Dorfman, J.~M.
  Hausdorff, {GaitAssist}: {A} wearable assistant for gait training and
  rehabilitation in {Parkinson}'s disease, in: 2014 {IEEE} {International}
  {Conference} on {Pervasive} {Computing} and {Communication} {Workshops}
  ({PERCOM} {WORKSHOPS}), 2014, pp. 135--137.
\newblock \href {http://dx.doi.org/10.1109/PerComW.2014.6815179}
  {\path{doi:10.1109/PerComW.2014.6815179}}.

\bibitem{tripoliti_automatic_2013}
E.~E. Tripoliti, A.~T. Tzallas, M.~G. Tsipouras, G.~Rigas, P.~Bougia,
  M.~Leontiou, S.~Konitsiotis, M.~Chondrogiorgi, S.~Tsouli, D.~I. Fotiadis,
  Automatic detection of freezing of gait events in patients with {Parkinson}'s
  disease, Comput Methods Programs Biomed 110~(1) (2013) 12--26.
\newblock \href {http://dx.doi.org/10.1016/j.cmpb.2012.10.016}
  {\path{doi:10.1016/j.cmpb.2012.10.016}}.

\bibitem{hammerla_preserving_2013}
N.~Y. Hammerla, R.~Kirkham, P.~Andras, T.~Ploetz, On {Preserving} {Statistical}
  {Characteristics} of {Accelerometry} {Data} {Using} {Their} {Empirical}
  {Cumulative} {Distribution}, in: Proceedings of the 2013 {International}
  {Symposium} on {Wearable} {Computers}, {ISWC} '13, ACM, New York, NY, USA,
  2013, pp. 65--68.
\newblock \href {http://dx.doi.org/10.1145/2493988.2494353}
  {\path{doi:10.1145/2493988.2494353}}.

\bibitem{azevedo_coste_detection_2014}
C.~Azevedo~Coste, B.~Sijobert, R.~Pissard-Gibollet, M.~Pasquier, B.~Espiau,
  C.~Geny, Detection of {Freezing} of {Gait} in {Parkinson} {Disease}:
  {Preliminary} {Results}, Sensors 14~(4) (2014) 6819--6827.
\newblock \href {http://dx.doi.org/10.3390/s140406819}
  {\path{doi:10.3390/s140406819}}.

\bibitem{rodriguez-martin_posture_nodate}
D.~RodrÃ­guez-MartÃ­n, A.~SamÃ , C.~PÃ©rez-LÃ³pez, A.~CatalÃ , J.~Cabestany,
  A.~RodrÃ­, Posture {Detection} with waist-worn {Accelerometer}: {An}
  application to improve {Freezing} of {Gait} detection in {Parkinson}â€™s
  disease patients  6.

\bibitem{zhao_feasibility_2016}
Y.~Zhao, J.~H. Nonnekes, E.~J.~M. Storcken, S.~Janssen, E.~E. H.~v. Wegen,
  B.~R. Bloem, L.~D.~A. Dorresteijn, J.~P. P.~v. Vugt, T.~Heida, R.~J. A.~v.
  Wezel, Feasibility of external rhythmic cueing with the {Google} {Glass} for
  improving gait in people with {Parkinson}â€™s disease, Journal of neurology
  263~(6) (2016) 1156--1165.

\bibitem{rodriguez-martin_home_2017}
D.~Rodriguez-Martin, A.~Sama, C.~Perez-Lopez, A.~Catala, J.~M.~M. Arostegui,
  J.~Cabestany, Ã.~Bayes, S.~Alcaine, B.~Mestre, A.~Prats, M.~C. Crespo, T.~J.
  Counihan, P.~Browne, L.~R. Quinlan, G.~OLaighin, D.~Sweeney, H.~Lewy,
  J.~Azuri, G.~Vainstein, R.~Annicchiarico, A.~Costa, A.~RodrÃ­guez-Molinero,
  Home detection of freezing of gait using support vector machines through a
  single waist-worn triaxial accelerometer, PLOS ONE 12~(2) (2017) e0171764.
\newblock \href {http://dx.doi.org/10.1371/journal.pone.0171764}
  {\path{doi:10.1371/journal.pone.0171764}}.

\bibitem{murad_deep_2017}
A.~Murad, J.-Y. Pyun, Deep {Recurrent} {Neural} {Networks} for {Human}
  {Activity} {Recognition}, Sensors (Basel) 17~(11).
\newblock \href {http://dx.doi.org/10.3390/s17112556}
  {\path{doi:10.3390/s17112556}}.

\bibitem{camps_deep_2018}
J.~Camps, A.~SamÃ , M.~MartÃ­n, D.~RodrÃ­guez-MartÃ­n, C.~PÃ©rez-LÃ³pez, J.~M.
  Moreno~Arostegui, J.~Cabestany, A.~CatalÃ , S.~Alcaine, B.~Mestre, A.~Prats,
  M.~C. Crespo-Maraver, T.~J. Counihan, P.~Browne, L.~R. Quinlan, G.~Ã.
  Laighin, D.~Sweeney, H.~Lewy, G.~Vainstein, A.~Costa, R.~Annicchiarico,
  Ã.~BayÃ©s, A.~RodrÃ­guez-Molinero, Deep learning for freezing of gait
  detection in {Parkinson}â€™s disease patients in their homes using a
  waist-worn inertial measurement unit, Knowledge-Based Systems 139 (2018)
  119--131.
\newblock \href {http://dx.doi.org/10.1016/j.knosys.2017.10.017}
  {\path{doi:10.1016/j.knosys.2017.10.017}}.

\bibitem{hochreiter_long_1997}
S.~Hochreiter, J.~Schmidhuber, Long {Short}-{Term} {Memory}, Neural Comput.
  9~(8) (1997) 1735--1780.
\newblock \href {http://dx.doi.org/10.1162/neco.1997.9.8.1735}
  {\path{doi:10.1162/neco.1997.9.8.1735}}.

\bibitem{eck_first_2002}
D.~Eck, J.~Schmidhuber, A {First} {Look} at {Music} {Composition} {Using}
  {LSTM} {Recurrent} {Neural} {Networks}, Tech. rep., Istituto Dalle Molle Di
  Studi Sull Intelligenza Artificiale (2002).

\bibitem{graves_supervised_2012}
A.~Graves, Supervised {Sequence} {Labelling} with {Recurrent} {Neural}
  {Networks}, Vol. 385, 2012.

\bibitem{liwicki_novel_2007}
M.~Liwicki, A.~Graves, H.~Bunke, J.~Schmidhuber, A novel approach to on-line
  handwriting recognition based on bidirectional long short-term memory
  networks, in: In {Proceedings} of the 9th {International} {Conference} on
  {Document} {Analysis} and {Recognition}, {ICDAR} 2007, Springer, 2007.

\bibitem{graves_framewise_2005}
A.~Graves, J.~Schmidhuber, Framewise phoneme classification with bidirectional
  {LSTM} networks, in: Proc. {IEEE} {International} {Joint} {Conference} on
  {Neural} {Networks}, Vol.~4, 2005, pp. 2047--2052 vol. 4.
\newblock \href {http://dx.doi.org/10.1109/IJCNN.2005.1556215}
  {\path{doi:10.1109/IJCNN.2005.1556215}}.

\bibitem{graves_connectionist_2006}
A.~Graves, S.~FernÃ¡ndez, F.~Gomez, J.~Schmidhuber, Connectionist {Temporal}
  {Classification}: {Labelling} {Unsegmented} {Sequence} {Data} with
  {Recurrent} {Neural} {Networks}, in: Proceedings of the 23rd {International}
  {Conference} on {Machine} {Learning}, {ICML} '06, ACM, New York, NY, USA,
  2006, pp. 369--376.
\newblock \href {http://dx.doi.org/10.1145/1143844.1143891}
  {\path{doi:10.1145/1143844.1143891}}.

\bibitem{graves_hybrid_2013}
A.~Graves, N.~Jaitly, A.-r. Mohamed, Hybrid speech recognition with deep
  bidirectional {LSTM}, in: In {IEEE} {Workshop} on {Automatic} {Speech}
  {Recognition} and {Understanding} ({ASRU}, 2013.

\bibitem{graves_speech_2013}
A.~Graves, A.~r. Mohamed, G.~Hinton, Speech recognition with deep recurrent
  neural networks, in: 2013 {IEEE} {International} {Conference} on {Acoustics},
  {Speech} and {Signal} {Processing}, 2013, pp. 6645--6649.
\newblock \href {http://dx.doi.org/10.1109/ICASSP.2013.6638947}
  {\path{doi:10.1109/ICASSP.2013.6638947}}.

\bibitem{zhang_highway_2015}
Y.~Zhang, G.~Chen, D.~Yu, K.~Yao, S.~Khudanpur, J.~Glass, Highway {Long}
  {Short}-{Term} {Memory} {RNNs} for {Distant} {Speech} {Recognition},
  arXiv:1510.08983 [cs]ArXiv: 1510.08983.

\bibitem{inoue_deep_2016}
M.~Inoue, S.~Inoue, T.~Nishida, Deep {Recurrent} {Neural} {Network} for
  {Mobile} {Human} {Activity} {Recognition} with {High} {Throughput},
  arXiv:1611.03607 [cs]ArXiv: 1611.03607.

\bibitem{chollet2015keras}
F.~Chollet, et~al., Keras, \url{https://keras.io} (2015).

\bibitem{demsar2006statistical}
J.~Dem{\v{s}}ar, Statistical comparisons of classifiers over multiple data
  sets, The Journal of Machine Learning Research 7 (2006) 1--30.

\bibitem{APB:05b}
M.~Atzmueller, F.~Puppe, H.-P. Buscher, {Profiling Examiners using Intelligent
  Subgroup Mining}, in: Proc. 10th International Workshop on Intelligent Data
  Analysis in Medicine and Pharmacology (IDAMAP-2005), Aberdeen, Scotland,
  2005, pp. 46--51.

\bibitem{valmarska17}
A.~Valmarska, N.~Lavrac, J.~F\"urnkranz, M.~Robni-Sikonja, {Refinement and
  Selection Heuristics in Subgroup Discovery and Classification Rule Learning},
  Expeert Systems with Applications 81.

\bibitem{ABP:06a}
M.~Atzmueller, J.~Baumeister, F.~Puppe, {Semi-Automatic Learning of Simple
  Diagnostic Scores utilizing Complexity Measures}, Artificial Intelligence in
  Medicine 37~(1) (2006) 19--30.

\bibitem{PABHLB:08}
F.~Puppe, M.~Atzmueller, G.~Buscher, M.~Huettig, H.~LÃ¼hrs, H.-P. Buscher,
  {Application and Evaluation of a Medical Knowledge-System in Sonography
  (SonoConsult)}, in: Proc. 18th European Conference on Artificial Intelligence, 2008, pp. 683--687.

\bibitem{battaglia2018relational}
P.~W. Battaglia, J.~B. Hamrick, V.~Bapst, A.~Sanchez-Gonzalez, V.~Zambaldi,
  M.~Malinowski, A.~Tacchetti, D.~Raposo, A.~Santoro, R.~Faulkner, et~al.,
  {Relational Inductive Biases, Deep Learning, and Graph Networks},
  arXiv:1806.01261.

\bibitem{AHTK:18}
M.~Atzmueller, N.~Hayat, M.~Trojahn, D.~Kroll, {Explicative Human Activity
  Recognition using Adaptive Association Rule-Based Classification}, in: Proc.
  IEEE International Conference on Future IoT Technologies, IEEE, Boston, MA,
  USA, 2018.

\end{thebibliography}

\end{document}